# The Galactic Exoplanet Survey Telescope (GEST)

D. P. Bennett[a], J. Bally[b], I. Bond[c], E. Cheng[d], K. Cook[e], D. Deming[d], P. Garnavich[a], K. Griest[f], D. Jewitt[g], N. Kaiser[g], T. Lauer[h], J. Lunine[i], G. Luppino[g], J. Mather[d], D. Minniti[j], S. Peale[k], S. Rhie[a], J. Rhodes[d], J. Schneider[l], G. Sonneborn[d], R. Stevenson[a], C. Stubbs[m], D. Tenerelli[n], N. Woolf[i], and P. Yock[c]

[a] University of Notre Dame, Notre Dame, IN, USA
[b] University of Colorado, Bolder, CO, USA
[c] University of Auckland, Auckland, New Zealand
[d] NASA/Goddard Space Flight Center, Greenbelt, MD, USA
[e] Lawrence Livermore National Laboratory, Livermore, CA, USA
[f] University of California, San Diego, CA, USA
[g] University of Hawaii, Manoa, HI, USA
[h] National Optical Astronomy Observatories, Tucson, AZ, USA
[i] University of Arizona, Tucson, AZ, USA
[j] Universidad Catolica de Chile, Santiago, Chile
[k] University of California, Santa Barbara, CA, USA
[l] Observatoire de Paris, Meudon, France
[m] University of Washington, Seattle, WA, USA
[n] Lockheed Martin Space Systems Company, Sunnyvale, CA, USA

**ABSTRACT**

The Galactic Exoplanet Survey Telescope (GEST) will observe a 2 square degree field in the Galactic bulge to search for extra-solar planets using a gravitational lensing technique. This gravitational lensing technique is the only method employing currently available technology that can detect Earth-mass planets at high signal-to-noise, and can measure the frequency of terrestrial planets as a function of Galactic position. GEST's sensitivity extends down to the mass of Mars, and it can detect hundreds of terrestrial planets with semi-major axes ranging from 0.7 AU to infinity. GEST will be the first truly comprehensive survey of the Galaxy for planets like those in our own Solar System.

## 1. INTRODUCTION

The search for extra-solar planets is a major part of NASA's vision for the Twenty-first Century, and the gravitational microlensing planet search technique has an important role to play in this effort. The main goal behind NASA's extra-solar planet search effort is to learn about how life may have developed outside the Solar System. Thus, planets that are thought to be habitable are of greatest interest, particularly if they could be inhabited by intelligent life.

The issue of planetary habitability is a complex and poorly defined one. The Earth's habitability is a consequence of a complex interplay of physical processes[1] that are not likely to be replicated in exactly the same way on other worlds. While the fundamental requirement is stable liquid water over geologic time, many diverse factors come into play in establishing habitable ecosystems[2]. More importantly, we do not know what the outcome of a different combination or timing of such processes would be in terms of habitability[3]. A non-exhaustive list of the potential requirements for habitability include the presence of giant planets in ~5-10 AU orbits[4], the presence of a large moon to stabilize the planetary spin axis[5], and main sequence stellar type of F, G, or K[6,7]. Also, the traditional notion that a narrow range of semi-major axes are consistent with the presence of liquid water[8] is challenged by the evidence for liquid water on the early Mars[9]. The length and incompleteness of this shopping list demands survey missions be initiated soon to map out the geometries of extra-solar planetary systems prior to much more expensive missions whose intent is to spectroscopically examine extra-solar terrestrial planets. With its high sensitivity to low-mass planets at a wide range of separations, GEST is the ideal mission for a comprehensive survey of the properties of planetary systems.

The search for exoplanets to date is restricted to large masses and short orbital periods. Our notion of the solar system as the archetypal model of all solar systems has been severely challenged by recent detections of extrasolar planets. The radial velocity technique[9] has revealed gas giant planets orbiting about 5% of nearby stars with a mass function that is rising down to their sensitivity limit. Nearly half of these planets are well within the orbital radius of Mercury, and only one has an orbital period longer than 10 years. Many have highly eccentric orbits, but only one[10], 47 Ursa Majoris, resembles our solar system. The 95% of systems without detected planets might resemble the Solar System, or they could be very different. These recent observations suggest that orbital migration and strong mutual gravitational interactions between planets play important roles in the evolution of these planetary systems, and both of these processes would result in the ejection of a large number of planets during the formation process[11]. GEST is the only proposed method that can detect a population of free-floating planets that would result from these planetary ejections.

## 2. THE GRAVITATIONAL MICROLENSING TECHNIQUE

The gravitational microlensing technique is well understood and well proven. The physical basis of gravitational microlensing is the gravitational attraction of light rays by a massive body such as a star or a planet. As illustrated in Fig. 1, if a "lens star" passes very close to the line of sight to a more distant source star, the gravitational field of the foreground lens star will deflect the light rays from the source star. The gravitational focusing effect of the lens star shifts, "splits", distorts, and magnifies the images of the source star. With sufficient resolution, two images would be visible to the observer. For Galactic microlensing, the image separation is < 1 mas, so the observer sees a microlensing event as a transient brightening of the source star as the lens star's proper motion carries it across the line-of-sight.

The Einstein ring radius, $R_E$, characterizes a gravitational microlensing event,

$$R_E = 2.0\mathrm{AU}\sqrt{\frac{M_{\mathrm{lens}}}{0.5 M_\odot}\frac{D}{1\mathrm{kpc}}}$$

where $D$ is the reduced distance given by $1/D = 1/D_l + 1/D_s$, $D_l$ is the distance to the lens, and $D_s$ is the distance to the source. This is the radius of the ring image that is seen if the lens and source stars are perfectly aligned. The lensing magnification is determined by the alignment of the lens and source stars measured in units of $R_E$, so even low mass lenses can give rise to high magnification microlensing events. The duration of a microlensing event is given by the Einstein ring crossing time, which is typically a month or two for stellar lenses, and a few days or less for a free-floating planet.

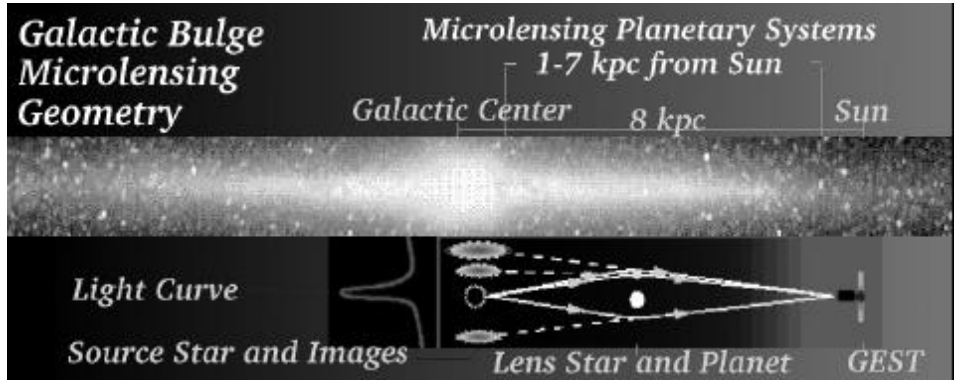

**Figure 1.** The geometry of a planetary microlensing event in the Galactic bulge. Bulge main sequence stars are monitored for magnification due to gravitational lensing by foreground stars and planets in the Galactic disk and bulge.

In 1993 the MACHO[12], EROS[13], and OGLE[14] groups discovered the first gravitational microlensing events, and by now, more than one thousand of microlensing events have been observed[15,16].

The microlensing signals of planets are distinctive and diagnostic. An extra-solar planet orbiting a lens star becomes detectable when the angular position of the planet comes very close to the angular position of one of the images due to the stellar lens. The planet's gravity further deflects the light rays of this image resulting in additional bright images and a deviation of the microlensing light curve from a normal single lens light curve[17]. As shown in Fig. 2, this gives a wide variety of different light curve shapes.

Microlensing is most sensitive to planets at a separation of ~ $R_E$ from the lens star. For stellar mass lenses and microlensing events towards the Galactic bulge, $R_E$ is typically 1-5 AU, so this is the region of maximum sensitivity for the microlensing planet search technique[18].

The most important feature of planetary microlensing is that these planetary deviations are large with typical variations of ~10%, even for planets of less than an Earth mass[19]. These signals are the strongest of any proposed Earth-mass planet search technique as was recognized by the "HST and Beyond Committee" who stated that microlensing is "a promising new way to determine the statistical occurrence of planets, including small planets like Earth."

Planetary light curves from microlensing are unique. The large photometric planetary signals that the microlensing technique offers would be of limited usefulness if there were other types of perturbations that could mimic microlensing perturbations. This issue has been studied in detail by Gaudi & Gould[20] and Gaudi[21] who found one type of microlensing perturbation that seemed capable of mimicking a planet detection. Most planetary microlensing light curves have positive and negative brightness deviations from the single lens light curve [see Figs. 2 (a), (b), and (d)], which cannot be mimicked by other non-planetary perturbations. But, if the source star is a member of a binary system with a large brightness ratio, then the secondary star could achieve near perfect alignment with the lens and undergo microlensing magnification that would be detectable on top of the microlensing light curve of the primary. However, the detailed shape of such a binary source light curve does not match the shape of a microlensing planet light curve. So, as long as we have complete light curve coverage for 2-3 days centered on the planetary deviation, the contamination of the planetary microlensing sample by non-planetary events is negligible.

Important planet parameters can be derived from microlensing. The utility of planet detections depends a great deal on what planetary parameters can be extracted from the detected planets. Planetary microlensing light curves can be classified in terms of the shape, duration, and the magnification due to the stellar lens at the time of the planetary deviation[18,19,22,23]. The modeling of a planetary microlensing light curve will determine the planetary mass fraction, $\varepsilon$, and the transverse separation of the planet from the lens star in units of the Einstein radius, $R_E$. As long as there is complete coverage of the planetary deviation region of the light curve, these parameters are accurately measured, except for a few percent of planetary events where there is some ambiguity in the determination of the separation[24]. In Section 2.2.1, we show that it will be possible to identify the planetary host star for one third of GEST's microlensing planetary discoveries. This will allow us to directly determine the stellar and planetary masses as well as the star-planet separation in AU for this subset of stars, which includes nearly all of the F, G, and K host stars.

Planets well below an Earth mass can be detected. Low-mass planetary signals are rarer and briefer than the microlensing signal from high-mass planets, so a larger number of stars must be observed more frequently to have high sensitivity to low-mass planets. The fundamental sensitivity limit occurs when the planet mass becomes too small to magnify more than a small fraction of a star at a time[19]. For giant source stars this limit is slightly above an Earth mass, but for main sequence source stars, this limit occurs at about the mass of Mars, so a low-mass planet search program must monitor main sequence source stars.

A unique capability of microlensing is its ability to detect free-floating planets. These can be detected down to a Mars mass as very short timescale microlensing events. Such planets are expected to be very common because the late stages of planetary system formation generally involve planetary scattering events in which a close approach between two planets results in one of the planets (usually the less massive one) being ejected from the planetary system. It is thought that giant planets like Jupiter will routinely eject a large number of planets in the terrestrial mass range[11] during the early phases of planetary accretion. An important consequence of the free-floating planet survey is that should Earth-like planets prove to be rare, this survey might provide an explanation. If each star ejected one Earth mass planet, GEST would detect about 20 such objects.

## 2. GEST MISSION SCIENCE

GEST will survey ≳100 million stars. The main challenge faced by the GEST mission and gravitational microlensing surveys, in general, is that the precise alignment between the source and lens stars is quite rare. The star fields with the highest gravitational microlensing probability are the fields very close to the Galactic Center, but even in the best field visible in optical wavelengths, the microlensing probability is only about $4\times10^{-6}$. (Formally, this is the probability that a source star is magnified by a factor of more than 1.34. This is often referred to as the microlensing optical depth, $\tau$.) Thus, microlensing surveys must observe millions of stars. Microlensing can only detect a planet if one of the images due to the stellar lens passes close to the planet. For Earth-like planets, the probability[19] of this is ~ 2%, so it is important to survey ~100 million stars in order to maintain high sensitivity to low-mass planets.

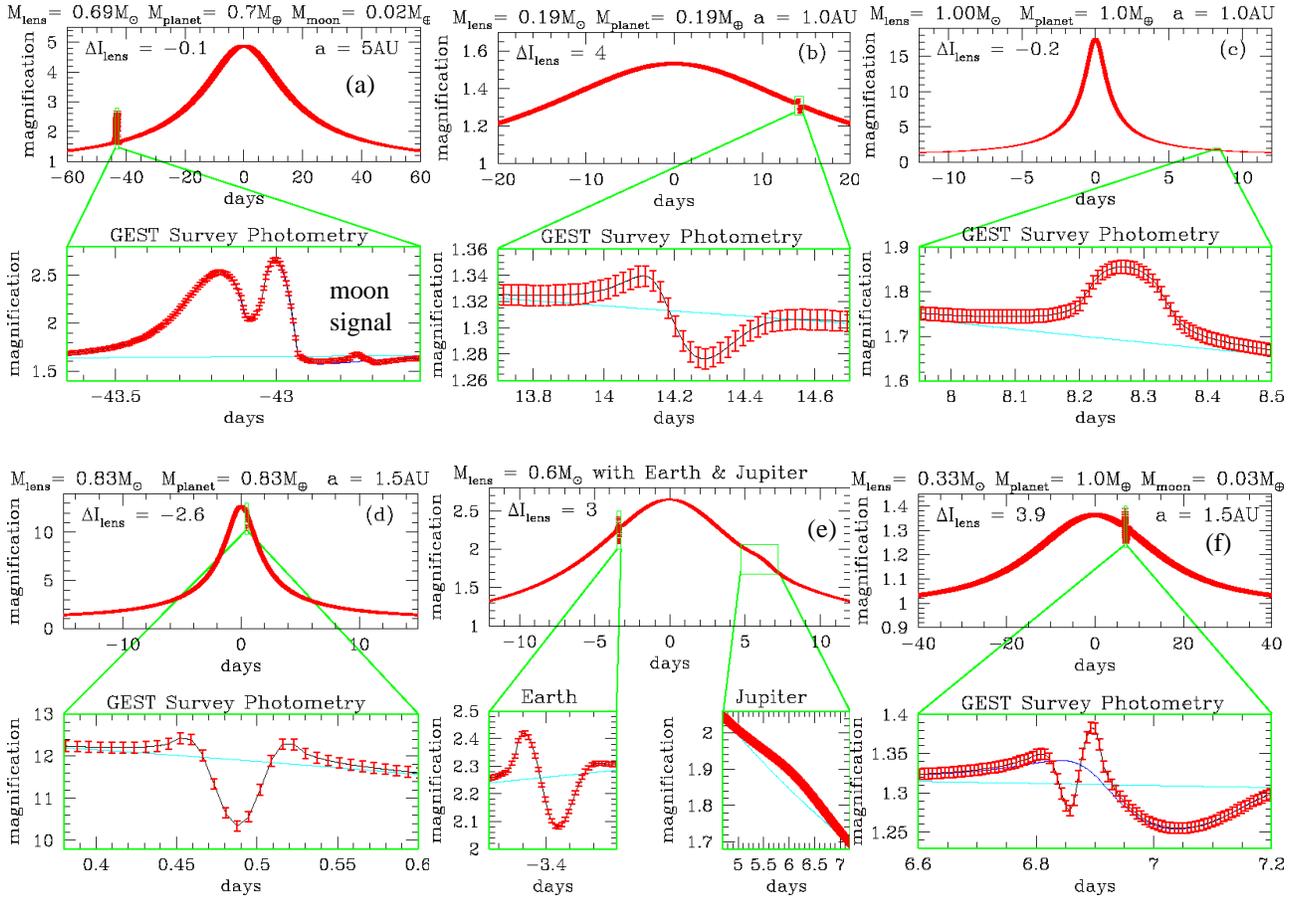

**Figure 2.** Example light curves from a GEST mission simulation. In each case, the top panel shows the full light curve, and the planetary deviation regions are blown up and shown in the lower panels. All of the example light curves have the Earth:Sun mass ration of $\varepsilon = 3 \times 10^{-6}$ except for Figure (f) which has $\varepsilon = 10^{-5}$. Figs. (a) and (b) span the range of planetary detection significance from $\Delta\chi^2 = 60,000$ (a) to $\Delta\chi^2 = 180$ (b) which is close to our cut. Figures (c) and (d) show more typical light curves with $\Delta\chi^2 = 600\text{-}1300$. Fig. (e) shows a multiple planet detection while Figs. (f) and (a) show light curve deviations due to moons of ~2 Lunar masses with a separation similar to that of the Earth's moon. $\Delta I_{lens}$ is the difference between lens and source I magnitude.

We are led to a unique choice for the star field for our planet search program-one that maximizes both the surface density of source stars and the microlensing probability. This is a field located at Galactic longitude and latitude l = 1, b = -2.5º, in the Galactic bulge. In this direction, GEST can view $\gtrsim 10^8$ main sequence stars brighter than I = 25 in a 2 square degree field-of-view, and the microlensing probability ranges[15] from 2.7-4×$10^{-6}$. The lens stars reside in the Galactic disk or bulge as shown in Fig. 1. They are a random sample of the stars along the line-of-sight, so they are predominantly M, K, G, and F stars.

## 2.1 The GEST Exoplanet Search Program

The GEST Baseline Mission runs 3.7 years. The GEST baseline mission includes four 8-month seasons of continuous observations of a dense star field near the Galactic Center to search for planets via gravitational microlensing. The remaining observing time is devoted a search for Kuiper Belt Objects and a Participating Scientist Program. The exoplanet search program requires that $\gtrsim$100 million main sequence stars in the Galactic bulge be observed continuously, so GEST must always be able to view the Galactic bulge. Also, the large number of stars that must be observed implies that a very large format detector must be used. This implies a very high data rate, so we require a nearly circular, geosynchronous orbit, which is highly inclined (> 45º) with respect to the ecliptic plane to maintain a continuous view of both the Galactic bulge and a dedicated ground station.

GEST requires modest photometric stability and very good pointing stability. The GEST instrument must provide near diffraction limited images (FWHM < 0.25" at 800 nm) over a field-of-view covering at least 2 square degrees in order to provide stable photometry of the large number of stars in the crowded fields observed for GEST's exoplanet search project. Our planetary sensitivity is maximized if we can attain a photometric sensitivity and stability of 0.3% for the brightest bulge main sequence stars, although most of our planet detections can be made with a 1% requirement. This requirement implies that we must map out the CCD sensitivity variations on sub-pixel scales[25], and this leads a stringent pointing stability requirement. The spacecraft must maintain pointing stability and drift to within 0.025" for at least 95% of the time during the Galactic bulge observing season. The telescope drift will be stabilized via a feedback loop using a signal generated by fast readout guide CCDs in the corners of the focal plane. This fine guidance system will enable GEST to execute a square 4×4 dither pattern consisting of 0.054" steps to allow interpolation across pixel boundaries. Dither steps are taken at intervals of 10-30 minutes.

Details of the instrument design are given in Section 2.1.1. We plan a CCD focal plane array of $6.0 \times 10^8$ pixels, which will result in images of 8.5 Gbits each assuming digitization at 14 bits per pixel. Images will generally be taken at 2-minute intervals, and the CCDs will be read out in less than 10 seconds while they are covered by a narrow shutter that will cover only one row of CCDs at a time. These 2-minute exposures will be co-added on board to yield combined 10-minute exposures. Cosmic rays will be removed with a simple algorithm such as median filtering and the data will be compressed using the Rice algorithm to reduce the data rate to 10 Mbits/sec or less and transmitted to our dedicated North American ground station in the X-band. This can be accomplished with one flight qualifiable Actel FPGA per amplifier. When the data arrive at our ground station they will be immediately processed to photometric measurements by a dedicated parallel-computing data processing system using an image differencing algorithm (see Section 3.2.3.).

A microlensing analysis will be run automatically on the new microlensing data as they come in so that planetary events can be discovered while they are in progress. This will enable follow-up observations while the planetary signals are in progress. The microlensing event photometric time-series database will be the primary scientific output of the GEST exoplanet search program. The GEST science team will perform gravitational microlensing analyses on these data to determine the parameters of the newly discovered planets as well as statistical information on the abundance of planets as a function of their mass and separation. This analysis and the associated data will be published as soon as possible, and it is anticipated that the first planetary discoveries from the GEST mission will appear in print within a year after launch.

The GEST results will provide a revolutionary advance in our understanding of extra-solar planetary systems. GEST will measure the following for stars in the inner galaxy:

- The average number of planets per star with sensitivity that extends down to about the mass of Mars (one-tenth of an Earth mass!). *This is an improvement of three orders of magnitude beyond previous and existing searches.*
- The planetary mass function, $f(\varepsilon = M_p/M_*)$, which gives the abundance of planets as a function of the ratio of their mass to that of the host star.
- The abundance of free-floating planets with masses down to that of Mars.
- The ratio of the number of free-floating planets to the number of bound planets.
- The mean number of planets per star as a function of the stellar mass and Galactocentric distance for $0.3 M_\odot \leq M_* \leq 1.1 M_\odot$. This is the mass range where most of the lens stars can be identified.

## 2.2 GEST Simulation

A detailed simulation of the GEST mission has already been accomplished[26]. In order to determine GEST's planetary detection potential, we have carried out a detailed simulation of the GEST mission assuming that a field of 2 square degrees centered at Galactic coordinates: l = 1º, b = -2.5º, is monitored continuously for 4 seasons of 8 months duration. This simulation is based upon the following assumptions:

1. The source stars follow the luminosity function of Holtzman et al.[27], but with a higher star density and higher reddening which is appropriate for our more central field. The star density is assumed to range from 1.55 to 2.06 times the star density in Baade's Window (as determined by the relative number of red clump giant stars seen in the MACHO database), and the extinction is assumed to be $A_I$ = 1.6 magnitudes.
2. The lens stars are assumed to follow the Kroupa[28] mass function and the mass-luminosity relation of Kroupa & Tout[29].
3. The microlensing probability (or optical depth) $\tau$ is assumed to range from $\tau = 2.7 \times 10^{-6}$ to $\tau = 4 \times 10^{-6}$ for the outer and inner halves of the GEST survey field. These values are based upon recent results from the MACHO Project[25] scaled to account for the proximity of the GEST field to the Galactic center.
4. The GEST camera is assumed to detect 20 photons per second from an I = 22 star as expected for the predicted GEST instrument through-put using a 600-1000 nm passband and QE curves for the Lincoln Labs CCDs planned for the GEST mission.
5. GEST's photometric precision is assumed to be determined by photon statistics for noise levels down to 0.3% as has been demonstrated with HST images[25,30]. The photon noise includes contributions from neighboring stars, which may be blended with the source, and the lens star itself, which can sometimes outshine the source star.
6. Microlensing light curves, complete with noise estimates, are generated for a grid of planetary mass fractions, $\varepsilon$, and separations with random orbital inclinations and phases.

The Simulation confirms GEST's enormous sensitivity compared to other missions. The results of these simulations are summarized in Fig. 4 which shows the sensitivity of the GEST survey as a function of planetary mass fraction, $\varepsilon$, and separation. The shaded regions in this plot indicate the sensitivities of the ongoing radial velocity surveys, the planned Keck astrometry program, the Space Interferometry Mission (SIM) and the Kepler transit search mission. The parameters of a number of the known extra-solar planets are indicated, as are the parameters of the planets of our own solar system.

GEST is quite clearly the most sensitive mission for all semi-major axes >0.7AU, and it has a much broader range of sensitivity than the other exoplanet search techniques. Kepler is the only other mission that can find true Earth analogs around a significant number of stars, but Kepler's greatest sensitivity is to planets with smaller semi-major axes. GEST is sensitive to terrestrial planets at separations ≥ 0.7AU, and is therefore complementary to Kepler. GEST is also directly sensitive to the planetary mass while Kepler provides an estimate of the planetary radius.

As Fig. 3 indicates, GEST can probe a full order of magnitude lower in mass than SIM. This figure displays GEST's planet detection sensitivity in terms of the planetary mass fraction, $\varepsilon$, which is the parameter most directly measured from a planetary microlensing light curve. Since the typical lens star is less massive than the Sun, the typical mass fraction of an Earth mass planet is close to $\varepsilon = 10^{-5}$. The fractional mass of the Earth to the Sun is $\varepsilon = 3 \times 10^{-6}$.

GEST's signal-to-noise for planet detection is high. A very attractive feature of the microlensing planet search technique is that low mass planets can give strong signals. Our detection threshold is set to be $\Delta\chi^2 \geq 160$ where $\Delta\chi^2$ is the $\chi^2$ difference between the best single lens microlensing event fit and the best planetary lensing fit. This is equivalent to a 12.5σ detection, most events signals that at >25σ.

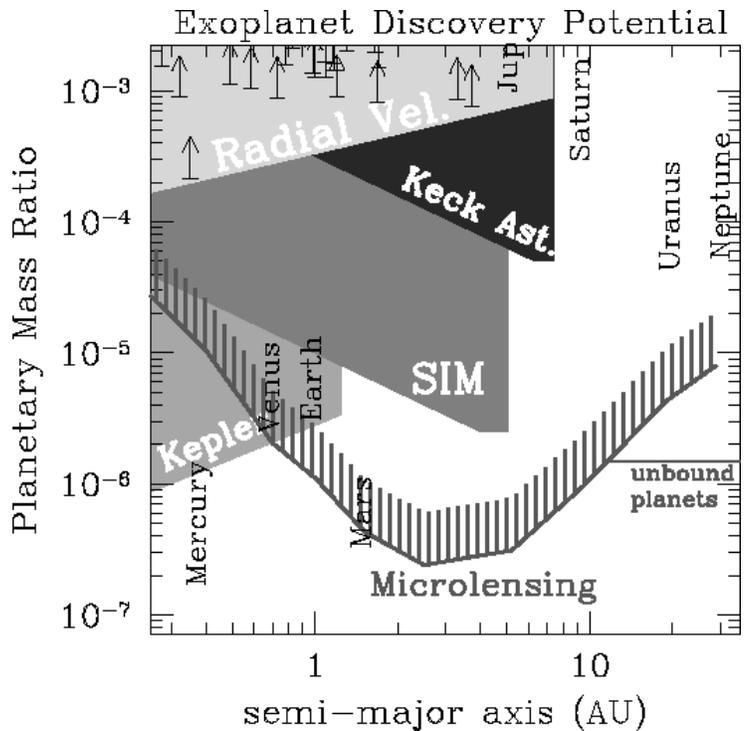

**Figure 3.** The sensitivities of the leading planet search techniques are plotted as a function of planetary mass fraction, $\varepsilon$. GEST will survey at least 10 systems for planetary systems with parameters above the curve labeled "Microlensing". The shaded regions indicate the parameter space that can be searched for planets via the radial velocity method, the planned Keck astrometry program, and the planned SIM and Kepler missions. The location of our Solar System's planets is labeled, and the arrows indicate the locations of known exoplanets.

It is apparent from Fig. 2 that there are a wide variety of light curve deviations that can be caused by planets. This is useful because it makes it relatively easy to measure the planetary mass ratio, ε, and the planetary separation from the microlensing light curves (measured in units of the Einstein ring radius which is typically 2-3 AU). It is also noteworthy that almost all of the planetary microlensing signals involve both positive and negative deviations from the normal stellar lens light curve. This is useful because the other sources of microlensing light curve deviations, such as binary source stars with a large brightness ratio, cannot produce light curve deviations with these shapes[21]. Thus, the relatively high signal-to-noise of the microlensing planet detections allows us to distinguish planetary microlensing from other types of light curve variations and to accurately determine the planetary parameters.

GEST is the only viable approach for detecting free-floating planets. In Fig. 3, it appears that planetary sensitivity drops off rapidly as the separation grows larger than ~10 AU. This is partly an artifact of our simulations, however. For these simulations, we have assumed that each planet is detected in conjunction with its parent star. However, for separations >>10 AU, it becomes increasingly likely that either the planet or the star will be detected, but not both. Photometric microlensing is relatively a short-range phenomenon, and a planet with a large semi-major axis is detected as a single lens event of a short duration. Completely unbound free-floating planets can be detected in the same way, and the duration will be 1-2 days for a Jupiter mass planet and a few hours for an Earth mass planet. Standard scenarios of planet formation predict a large population of free-floating planets, and gravitational microlensing is probably the only way to test this prediction. In fact, the MACHO Collaboration has observed[31] an event that appears to be caused by an object of about a Jupiter mass, but the sampling of the light curve is too sparse to be certain about this interpretation.

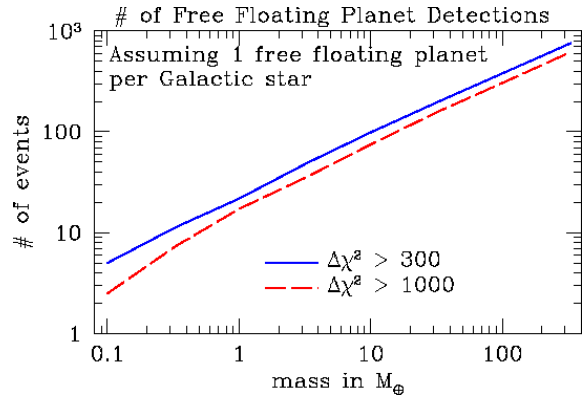

**Figure 4**. The number of free-floating planets to be discovered by GEST for 2 different detection criteria, which are equivalent to 17σ and 30σ respectively.

Free-floating planets cannot be distinguished from planets in distant orbits individually, but a substantial fraction of the very distant planets will show evidence of their parent star, so a statistical determination of the isolated planet abundance should be possible.

Because the abundance of free-floating planets is unknown, we assume that there is one free-floating planet per Galactic star, and then determine the number of free-floating planet discoveries expected from the GEST survey. This is shown as a function of planetary mass in Fig. 4.

The detection threshold of the planetary signals has been set higher than that in the case of the star plus planet microlensing events. The search for these short duration planetary signals from ~100 million light curves will have a larger number of false alarms than the search for planetary deviations in the ~$10^4$ stellar microlensing event light curves. If there is one free-floating Earth-mass planet per Galactic star, Fig. 4 indicates that GEST would detect 20 free-floating earths. However, standard theories of planetary system formation suggest that many Mars-Earth mass planets may be scattered out of each proto-planetary system, so if these theories are correct, we should expect a significant number of free floating planet detections down to the mass of Mars.

### 2.2.1 Many of the planetary host stars can be observed, including those of Solar type.

The planets detected by the GEST microlensing survey orbit the lens stars in the foreground of the Galactic bulge source stars. The mass distribution of the lens stars from our GEST simulations is shown in Fig. 5. This distribution is somewhat flatter than the stellar mass function because we have assumed that the planetary mass distribution is proportional to the stellar mass distribution and more massive planets have a higher detection probability.

Although the microlensing does not require the detection of any light from the lens stars, a significant fraction of the microlensing events seen by GEST will have lens stars that are bright enough to be detected. Our simulations indicate that for ~17% of the detected planets, the planetary host (lens) star is brighter than the source star, and for another ~23% the lens stars that is within 2.5 I-band magnitudes of the source star's brightness. A few of these stars are blended with the images of brighter stars which do not participate in the lensing event, and this leaves 33% of the lens stars should be directly detectable. The detectable planetary host stars are depicted in black in the left panel of Fig. 5, and they comprise virtually all of the F and G star lenses, most of the K star lenses, and a few of the nearby M-star lenses.

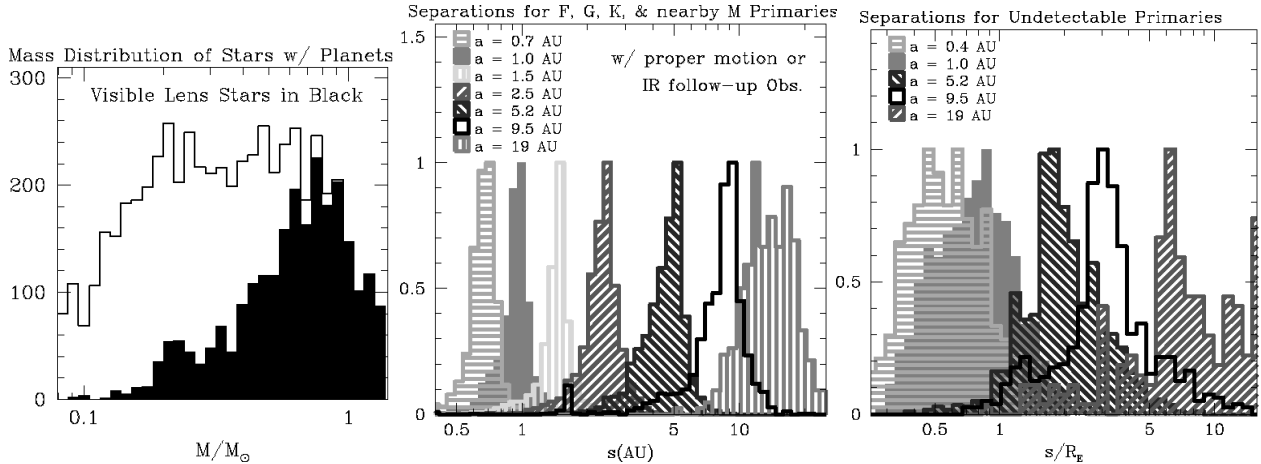

**Figure 5.** The simulated distribution of stellar masses is shown in the left panel for stars with detected terrestrial planets. Lens stars are considered visible when they are at least 10% of the brightness of the source star, if they are not blended with a brighter star (besides the source). 1/3 of the events have visible lens stars. For these stars, it is possible to determine the Einstein Ring radius, $R_E$, in physical units. This allows the conversion of the measured separation, s, into physical units (AU), and the center panel shows the relationship between the actual 3-d star-planet separations and the measured 2-d separations. The observed scatter in the measured separation relation is mostly due to the projection of the orbital plane on the sky. The right panel shows the distribution of measured star-planet separations for detected planets, which orbit undetectable stars.

The visibility of the lens star will allow for the measurement of a number of other useful parameters. The most obvious of these are the apparent magnitude and color of the lens star. This would enable an approximate determination of the lens mass and distance if the dust extinction was small. Our field, however, has high and variable extinction, and so it will be prudent to obtain IR photometry. This will allow us to estimate both the extinction and the intrinsic color of the star. Because our fields are quite crowded, we will need IR observations with high angular resolution which can be obtained with an IR camera on HST or with adaptive optics (AO) systems on large telescope such as the VLT, Gemini, LBT and Keck. Because our targets are in crowded fields, we are virtually guaranteed that guide stars will be available to provide the phase reference needed for these AO systems. We would expect to obtain two sets of IR AO observations: one during the event which would be scheduled as soon as the planetary signal is detected, and the second set of observations which would be taken well after the event is over. This pair of observations taken at different lens magnifications will allow us to unambiguously determine the color and brightness of the lens stars. We will require these data only for events with detected planetary signals, and we anticipate no difficulty in obtaining the ground-based telescope time since six team members have access to substantial telescope time on privately owned telescopes in Chile, Hawaii, and Arizona.

Another measurable parameter for the visible lens stars is the relative proper motion between the lens and the source, which is typically $\mu \approx 8$ mas/yr for a total motion of 28 mas or 13% of a CCD pixel. Anderson and King[32] argue that centroids can be measured to 0.2% of a pixel with a combination of multiple undersampled HST WFPC2 frames. GEST will provide >100 times more data than the most ambitious HST programs, which will allow numerous crosschecks to look for systematic errors in the centroid determinations. Thus, we expect that the centroids of the GEST stars can be determined at least as well as the centroids of the HST stars, so we expect to be able to measure the relative proper motion to an accuracy of a few percent. An independent measurement of the lens-source proper motion can be obtained for the events, which exhibit planetary lens caustic crossing features. These comprise somewhat more than 50% of the events in which terrestrial planets are detected, and they allow the ratio of the angular radius of the star to the angular Einstein radius, $\theta_E$, to be measured in the light curve fit. Since the source star angular radius can be estimated from its brightness and color, an estimate of $\theta_E$ can be obtained. The ratio of the angular Einstein radius to the lens-source proper motion is $\theta_E/\mu = t_E$, the Einstein radius crossing time which can also be measured from the light curve, and so these measurements of $\mu$ and $\theta_E$ give equivalent information.

The measurement of $\mu$ or $\theta_E$ allows us to use the following relation for the lens star mass,

$$M_l = \frac{\theta_E^2 D_s c^2}{4G} \frac{x}{1-x}$$

where $x=D_l/D_s$, the ratio of the lens to source distances. This relation allows us to determine the difference between the source and lens distances when the lens is close to the source because it indicates that $M_l$ and hence the lens luminosity depends sensitively on $1-x = (D_s-D_l)/D_s$. This means that the Einstein radius, $R_E$, can be determined for all lens stars with a measurement of the lens star brightness and its relative proper motion, $\mu$, or its color. This means that the planetary separation can be determined in physical units. The results of this determination are shown in the central panel of Fig. 5, which shows the measured separation for detected planets as a function of their orbital semi-major axis. For this plot we have assumed that the change in the relative lens-source centroid can be measured to 2 mas, the reddening corrected I magnitude of the lens can be measured to an accuracy of 0.2 mag., and the reddening corrected I-K color can be measured to 0.1 mag. The resulting estimate for planetary semi-major axis is accurate to about 20% with the uncertainty dominated by the unmeasured component of the star-planet separation along the line-of-sight.

When the lens star cannot be detected, the projected separation between the planet and its host star can only be measured in units of $R_E$. This can be used to estimate the planetary orbit semi-major axis by means of the expected correlation shown in the right panel of Fig. 5, which indicates that physical separation can be estimated with an accuracy of a factor of 2 or 3.

### 2.2.2 Number of Expected Planetary Detections

Because GEST will search for planets that occupy an unexplored region of parameter space, it is impossible to accurately predict the number of planetary discoveries that GEST will make. However, if we present the number of expected planetary discoveries for an assumed distribution of planetary systems, we can gain a better understanding of GEST's sensitivity. In order to match the parameters used for our simulations, we will consider planetary mass fractions of $\varepsilon = 3\times10^{-7}$, $10^{-6}$, $3\times10^{-6}$, $10^{-5}$, $3\times10^{-5}$, $10^{-4}$, $3\times10^{-4}$, and $10^{-3}$ as well as separations of 0.4, 0.7, 1.0, 1.5, 2.5, 5.0, 9.5, 19, and 29 AU. Extrapolations based upon the extra-solar planets discovered by the radial velocity technique suggest that the number of planets per logarithmic mass interval is constant or increasing toward lower masses[33,34]. We will assume that there is a constant probability $f$ for a planet with each of our selected mass fractions, $\varepsilon$, to orbit each of our lens stars at each of the selected separations. This implies that the exoplanet mass function is constant with respect to $\log M$, and nearly independent of the logarithm of the separation. The average number of planets per system is $72f$ since we consider 8 possible planetary mass fractions and 9 different orbital separations. If we take $f = 0.1$, then we have an average of 7.2 planets per system, and our simulations show that GEST would detect 3000 planets. 130 of these would be in the terrestrial mass range ($\varepsilon \leq 10^{-5}$). The Kepler mission assumes[35] an average of 2 Earth-like planets ($\varepsilon = 3\times10^{-6}$ or $10^{-5}$) with separations between 0.5 and 1.5AU, and this corresponds to $f = 0.333$. With this value, GEST would see 430 terrestrial planets, and 280 of these would be inner orbit Earth-like planets with $\varepsilon = 3\times10^{-6}$ or $10^{-5}$ and separations $\leq$ 2.5AU. If we assume that there are as many free-floating as gravitationally bound planets, then GEST would detect 1400 free-floating planets, and 80 of these would be terrestrial planets.

### 2.3 Ground Based Microlensing Surveys cannot do GEST's mission

Gravitational microlensing has developed as a ground based observational technique over the past decade. The EROS, MACHO, OGLE, and MOA microlensing survey teams have now observed >1000 microlensing events caused by ordinary stars along the line of sight towards the Galactic bulge[15,16,36] and one case of microlensing by a probable free-floating gas giant planet[31]. This event has a much shorter timescale than the other detected events, and the source star is likely to be a giant with a radius of $\sim 10R_\odot$. With a peak magnification of ~5, differential magnification of the stellar disk should have produced a broad light curve maximum, which could have established the free-floating planet interpretation this event. But, the MACHO Project's nightly light curve sampling was too infrequent to see this effect

There are several strategies that can be used to obtain the high observing frequency needed for microlensing planetary discoveries. The PLANET[37] and MPS[38] Collaborations attempt to follow ongoing microlensing events with hourly sampling using networks of telescopes spanning the globe in the Southern Hemisphere. They rely upon dedicated microlensing survey experiments, such as MACHO[39], MOA[36] and OGLE[40] to announce the location of microlensing events in progress. To date, these groups have discovered 2 possible planetary microlensing events[41,42], but the interpretation of these events remains ambiguous[43,44] due to poor light curve coverage and poor observing conditions at critical times. Uneven light curve coverage does not prevent upper limits on the planetary abundance, and the PLANET

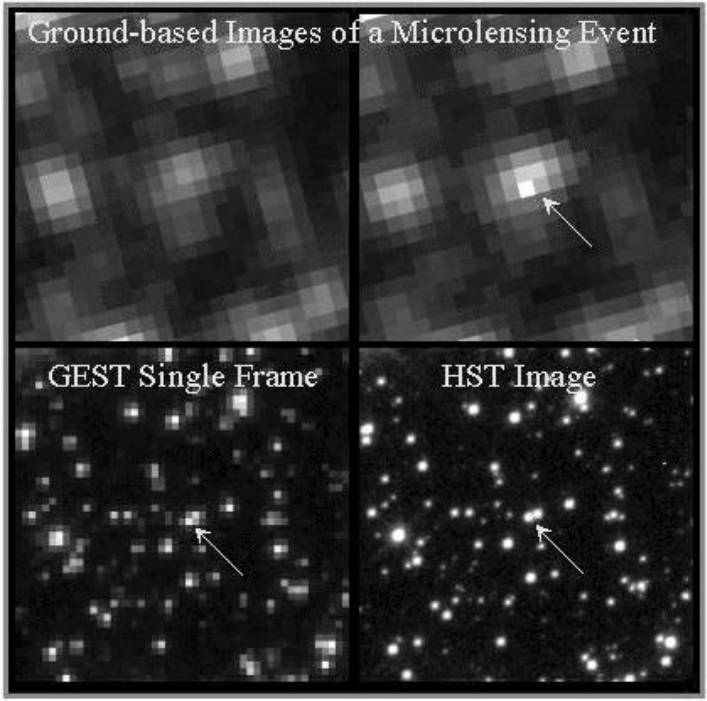

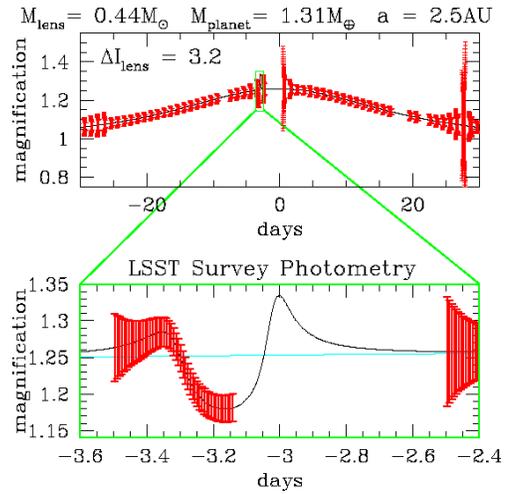

**Figure 7.** Light curve for a typical planetary deviation that can be detected by a ground based survey.

**Figure 6.** The difference between ground and space-based data for microlensing of a bulge main sequence star is illustrated with images of microlensing event MACHO-96-BLG-5. The two top panels are 50 min. R-band exposures with the CTIO 0.9m telescope, while the bottom images have been constructed from HST frames. The bottom left image represents a 10 minute exposure with GEST's angular resolution and pixel size, and the image on the right is a slightly degraded HST image, which represents the diffraction, limited image that can be reconstructed from the GEST dither pattern. The lensing magnification factors are A = 4 and 10 for the ground based images and 1.07 for the space based image. The source star, a Galactic bulge G-dwarf is indicated by the arrows.

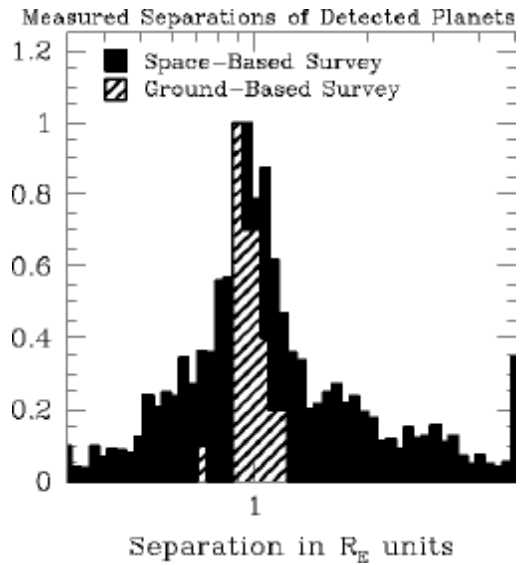

**Figure 8.** The separation distributions of detected planets for space-based and ground-based microlensing surveys are shown based upon realistic simulations these surveys. Note that the ground-based surveys also detect and characterize about 100 times fewer planets than a space-based survey.

team has obtained the strongest limits to date on giant planets in Jupiter-like orbits[45].

A much more ambitious microlensing planet search and follow-up program was proposed by Tytler et al[46] who advocated a Southern Hemisphere network of new 2-m class telescopes to follow microlensing events around the clock in attempt to detect Earth-mass planets. Peale[47] has studied the sensitivity of such a survey using realistic weather patterns from Southern Hemisphere observing sites. He found that only a few Earth-like planets could be detected in an 8-year survey, even though his estimate did not include the adverse effects of seeing variations or poor seeing in Australia. Peale did not consider how well the mass and separation could be determined from the ground-based data.

There is an even more serious flaw with this proposed observing program. Ground based images (see Fig. 6) seem to indicate that the majority of visible stars in the Galactic bulge are "turn-off" stars which have just left the main sequence and are 1-2 magnitudes brighter than the brightest main sequence stars, and it is these "turn-off" stars that Tytler et al. proposed to monitor. However, as Fig. 6 clearly shows, these apparent turn-off stars are actually blends of multiple unresolved main sequence stars at a density of a few per square arc second. When this stellar blending is taken into account, the signal-to-noise of the measurements drops significantly.

Blending difficulty also affects another type of ground based microlensing planet search program that was suggested by Sackett[48]. The typical Einstein radius crossing time for an Earth-mass planet is a few hours, so Sackett suggested that a wide field telescope observing from an excellent observing site, like Paranal, might be able to carry out a survey for Earth-mass planets. Such a program has been suggested for the VLT Survey Telescope (VST) and the VISTA telescope.

A crucial difficulty with this idea is that the light curve deviation due to a planet must be sampled for ~5-10 Einstein radius crossing times in order to fully characterize the event. Such a ground based microlensing program can be simulated with calculations similar to the Bennett and Rhie[26] simulation of a space-based microlensing planet search program, but for ground-based observations, we must include seeing, cloud cover, sky brightness, and airmass variations. The sky brightness dependence on the phase of the moon and the airmass are relatively straightforward to calculate[49]. Seeing data from Paranal are available at the VLT web site. These data were used for realistic simulations of a ground based planet search from the VST, VISTA and LSST telescopes assumed to be observing from Paranal. An example light curves from this simulation is shown in Fig. 7. The event shown is a typical event with a detectable planetary signal, but the signal is too poorly sampled to characterize the deviation. Because higher mass planets are easier to detect, there will be many more poorly sampled signals from massive planets, and this will make low-mass planetary signals more difficult to identify.

Because of the severe crowding in these fields the source stars are usually blended with other stars, and so the photometric signal-to-noise is low. This means that long events will be preferentially detected in such a survey because they will have many more observations that indicate the planetary deviation, so there is a selection effect, which favors the detection of incomplete planetary signals. Nevertheless, there are still a few terrestrial planet events that can be detected and characterized by a VST survey if it runs for many years. These events tend to have very high magnification and a star-planet separation that is very close to the Einstein ring radius. The high magnification alleviates the blending problems with ground-based photometry, and some of the planetary deviations seen at high magnification have a duration that is much smaller than the typical duration for a planetary deviation. This means that a ground-based survey would probe a much smaller range of separations than a space-based survey would, as shown in Fig. 8.

Extensive simulations of realistic ground based surveys[50] indicate that the number of terrestrial planets detected and characterized will be a factor of 50-100 smaller than the number of terrestrial planets detected and characterized[26] by a space mission like GEST. Furthermore, as shown in Fig. 8, a ground based survey will only provide a measure of the planetary abundance at a separation very close to the Einstein Ring radius, while a space-based mission like GEST can provide the planetary abundance as a function of separation.

One might hope to improve the sensitivity of the ground-based, terrestrial planet searches by adding powerful new wide FOV telescopes in South Africa and Australia. However, the seeing and weather at these locations is substantially worse than in Chile, and so it would not be possible to replicate the quality of the light curves from a telescope in Chile. In fact, all potential observing sites in Australia have such poor seeing that it would be difficult to get a review committee to even consider putting an new expensive telescope there. So, the option of building new wide FOV telescopes in South Africa and Australia would be an ineffective way to complete the partial microlensing light curves that could be obtained from a single Chilean site, and the modest light curve improvement to be expected would not justify the cost of these new telescopes.

The preceding discussion indicates that ground-based microlensing planet searches can detect some terrestrial planetary signals, but their light curve coverage is too poor to make more than a few, if any, definitive discoveries of terrestrial exoplanets. This conclusion is borne out by the results from current ground-based microlensing planet search projects. The MPS and MOA groups have proposed two candidate planetary events[41,42], but subsequent studies[43,44] have indicated that the proposed planetary models are not unique. In both cases, the ambiguity is due to gaps in the data caused by bad weather or poor observing conditions. The main result of the PLANET Collaboration[45] is a limit on the abundance of giant planets from 5 years of observations with their global network of telescopes. Less than one third of stars are found to have Jupiter mass planets between 1.5 and 4 AU. The hypothetical LSST, VISTA, or VST surveys discussed above could, in principle, set upper limits on terrestrial planet abundance, but *only if* they detected no planetary signals. If terrestrial planets are detected, the data will not be sufficient to determine planetary parameters. A much more viable strategy for ground based planet search programs is to observe microlensing events with giant source stars and very high magnification events with main sequence source stars. Such a program could discover exoplanets in the gas and ice-giant mass range or put stringent limits on their abundance.

## 3. MISSION DESIGN

The GEST observatory will be in a nearly circular geosynchronous orbit inclined at 28.7° to the Equator and by about 52° to the Ecliptic plane as shown in Fig. 9. This allows continuous viewing of the Galactic Bulge Microlensing Survey (GBMS) field with an Earth avoidance angle of >40° in the March-to-October time frame as well as a continuous data downlink to a ground station in the US. During the rest of the year, telescope observations will be made primarily in the opposite continuous view zone. Science data will be continuously transmitted to a dedicated ground station in the US at a rate of < 10 Mbits/sec. Communications support from the TDRSS or DSN systems is not required except in emergencies.

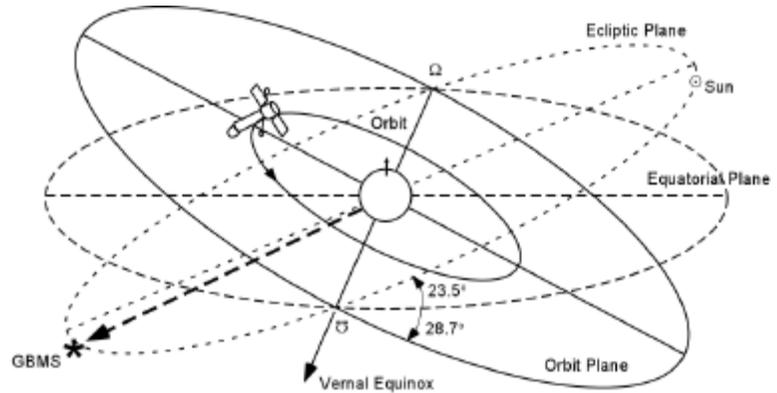

### 3.1 Sky Coverage

Fig. 9 depicts GEST's orbit and viewing constraints. The Galactic bulge microlensing survey field (GBMS) resides within one of the large continuous viewing zones (CVZ). With a 60° sun avoidance constraint, the only impediments to continuous access to the GBMS field during the 8-month March-to-October viewing season dictated by the sun's position is the once-a-month passage of the bright moon, which may interfere with observing for less than a day per month with little effect on the science.

**Figure 9.** GEST will fly in nearly circular, geosynchronous orbit inclined at 28.7° to the Equator and by about 52° to the Ecliptic plane.

Most of the GEST observations will be made in one of the two CVZs. During the November-to-February period when the solar constraint precludes GBMS viewing, second-priority programs to look for Kuiper Belt Objects, weak gravitational lensing, and high redshift supernovae will be implemented. Six to eight months will also be devoted to a guest observer program.

### 3.2 Instrument

The GEST instrument design utilizes existing, proven hardware and technologies. The instrument consists of a 1.0-1.5 m telescope, a focal plane assembly containing 32 backside thinned CCDs, and electronics. The GEST telescope structure and optics will be provided by COI and the University of Arizona, or else by Kodak. These are both low cost, low risk solution. The GEST Focal Plane Assembly (FPA) and electronics design is provided by Lincoln Labs, who have built a 10-CCD FPA for the ACIS instrument on Chandra.

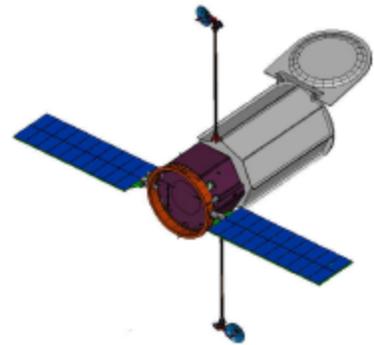

Accurate line of site pointing stability is required to allow the photometry software to compensate for the photometric effects of sub-pixel quantum efficiency variations of the CCDs. We will accomplish this with a closed-loop fine attitude control system to control low frequency S/C drift similar to the approach of our French collaborators on the COROT mission. Low frequency drift errors will be corrected using correlation tracking with imagery obtained using the GEST instrument as a high-precision star tracker. In addition to the science CCDs, 4 smaller guide CCDs are included in the focal plane assembly. These CCDs are read out 10 times a second and used for closed-loop instrument and spacecraft pointing.

**Figure 10.** The GEST telescope and spacecraft assembly.

The LM900 spacecraft, used for Ikonos in low Earth orbit, has demonstrated a mid- to high-frequency error of only 0.06 arc seconds. After adjusting for the increased inertia of the GEST observatory (about 8:1 greater than Ikonos), the GEST pointing jitter is expected to be less than 0.008 arc seconds. (GEST will have much smaller thermal disturbances than Ikonos due to its geosynchronous orbit.)

### 3.2.1 Telescope

The GEST telescope is a three mirror anastigmatic design with a 1.2×2.4 degree field-of-view. This design was selected to provide the wide field of view required by the scientific objectives, and it provides good baffling and a convenient pupil plane where filters can be mounted. The telescope structure and mirrors will be fabricated by COI and the University of Arizona or Kodak. A ray-trace drawing of the telescope optical design is shown in Fig. 11.

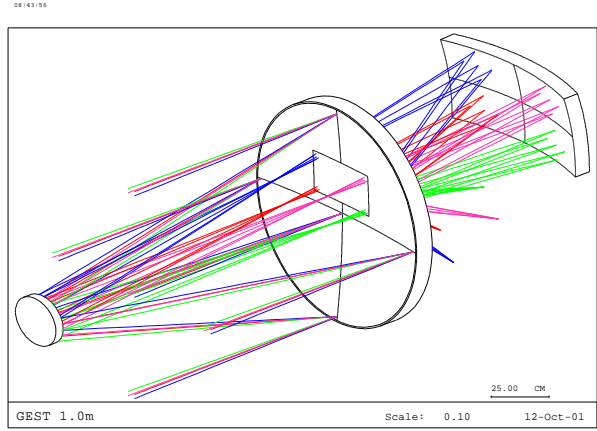

**Figure 11.** Optical design ray-trace for a 1.0m version of the GEST telescope design.

### 3.2.2 Focal Plane Assembly

The GEST Focal Plane Assembly (FPA) will include 32 large-scale (3k×6k-pixels), high quantum-efficiency (>80% at 800 nm), low-noise (~5 e$^-$) CCD devices. The devices will be cooled to approximately –90 C to reduce dark current to a level below the low-noise of the readout amplifiers. The FPA will therefore need to possess mechanical stability both to meet these requirements and to survive launch loads.

In order to meet the low-noise (~5 e$^-$) readout requirement and to minimize radiation damage effects, four readout amplifiers are needed for each CCD. This would normally require 144 channels of readout electronics, but it is feasible to share a 24-channel electronics package among 6 groups of CCDs that are read out consecutively. In either case, close spacing of the electronics channels and the large number of parallel channels needed imply very tight integration of the electronics to the FPA housing so that the electronics do not degrade the inherently low-noise CCD devices.

Lincoln CCDs have the large format and high red sensitivity required by GEST. Lincoln has already designed and fabricated CCDs in the required 3k×6k, 10 micron pixel format and Co-I Stubbs has 10 of these devices in hand. Only minor changes are required to optimize the CCD mask set for the GEST application, and so we expect that Lincoln will be able to fabricate the GEST CCDs at a high yield as they have done in the past.

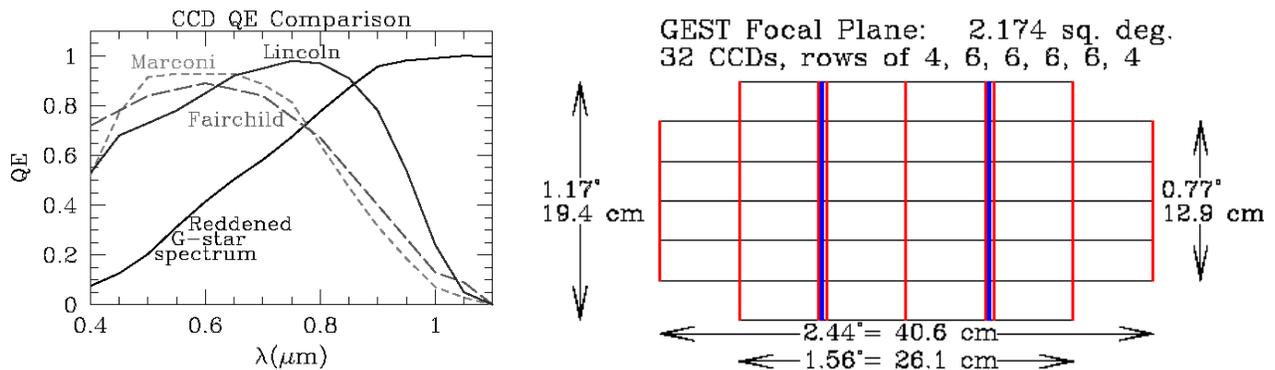

**Figure 12.** The left hand panel shows the quantum efficiency curve for the Lincoln Labs CCDs made from high resistivity silicon as compared to other space qualified CCDs, and the right panel shows the layout of the GEST focal plane.

Lincoln has developed a back-illuminated (BI) process, using very high-resistivity silicon substrates, which has the high quantum efficiency (QE) needed at red and near-IR wavelengths, as shown in Fig. 12. This process achieves a QE of nearly 80% at 900 nm, compared to 40% or less for more conventional BI CCDs. Lincoln has also measured the intra-pixel response of devices fabricated with this process and confirmed that the response is a smooth function of position across the pixel, as is required for accurate photometry.

Lincoln has built large FPAs for ground and space astronomy. Thirty-two CCDs will be assembled into a mechanically stable FPA that is capable of being cooled to –90° C. A sample layout for a 32-device FPA for GEST is shown in Fig. 12. The GEST FPA will require a new package and FPA technology that is capable of four-side abuttability, and such a package is being developed jointly between Lincoln and CoI Luppino.

Lincoln has previously demonstrated the ability to develop and launch successfully a large multi-device FPA with the 10-CCD ACIS instrument built by Lincoln for Chandra. This FPA was designed and tested at Lincoln to survive the mechanical and thermal loads predicted for the mission. Lincoln will carry out these tests for the GEST mission.

Lincoln will design and provide the multi-CCD FPA mounted inside a housing assembly. The housing will provide 2.5 cm of aluminum radiation shielding uniformly around the FPA. An optical window will be included with the dewar to provide for contamination control as well as radiation shielding. The proposed FPA design allows for accurate temperature control of the detectors with minimal complexity and power consumption. Temperature control is important to minimize relative shifts between pixels in the FPA, and also to optimize radiation hardness. In order to minimize thermal losses, the mounting of the FPA within the dewar will include low-thermal-conductivity standoffs, low-emissivity coatings on the inside of the dewar as well as the FPA structure, and low-thermal-conductivity electrical cables. The dewar will have cold fingers to provide an interface with thermal radiators provided by the spacecraft. Trim heaters and associated temperature-control sensors will be attached to the FPA for accurate detector temperature control. Temperature-control circuitry will be provided within the remotely located readout electronics.

GEST has a novel shutter design without a single point failure mode. The dewar assembly will also include mechanical shutters located above the FPA as shown in the conceptual drawing of Fig. 13. The shutters will be independently operable, driven by the remotely located readout electronics. One shutter will serve two adjacent rows of CCDs. The cross-section and placement of the shutters will be designed to minimize stray light reflections from the shutter blade and partial shadowing of edge pixels.

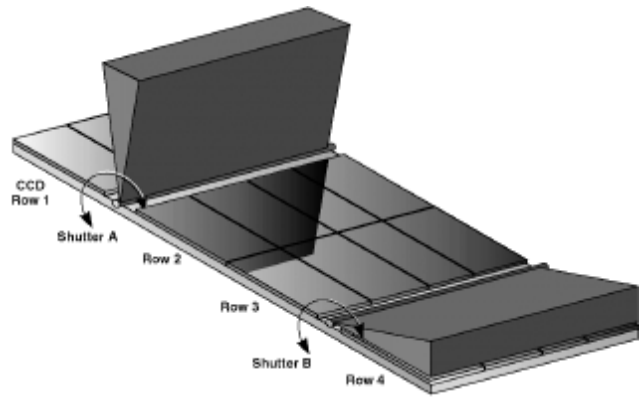

**Figure 13.** The GEST shutter concept.

The FPA contemplated for GEST will be a larger version of the one pictured in Fig. 13, and the device inputs and outputs will be distributed over the entire area of the FPA instead of only at the perimeter. This will require that the input and output of each device penetrate the bottom wall of the dewar. The close spacing of the I/O connectors requires that the electronics associated with each channel be compact, and located in a space below the bottom of the dewar. Lincoln has experience with similar electronics.

The compact electronics package developed for GEST will include timing generation and driver circuitry for the CCD circuits, and readout electronics that will provide 14-bit digital pixel values. We plan that one row of 6 devices, each with 4 outputs, which will be read out at a given time. The 24-channel readout electronics will be multiplexed among the 6 rows of devices. There will also be 4 small format, high-frame-rate CCD's at the corners of the FPA for spacecraft attitude control.

The GEST electronics package will provide on-board frame averaging and compression of the science CCD data. A frame-averaging control line will allow from one to ten frames to be averaged before the data are transferred to the spacecraft data recorder. Data compression by a factor >3.1 will be performed by a flight qualified Actel FPGA that can implement a version of lossless Rice encoding. This typically achieves a factor of 2-3 compression for astronomical images[51], and we anticipate that a substantial increase in the compression factor can be achieved with special purpose routines that make use of the fact that we are repeatedly imaging the same field. Another option is the square-root encoding scheme being considered for NGST[52], which achieves additional compression by limiting the resolution of the noise to 3 or 4 bits. This data compression module may be provided as a contribution to GEST by CNES.

The data from the 4 high-speed attitude-control CCDs will be converted into control signals and fed directly to the spacecraft attitude control system, bypassing the frame averaging system. The software to convert the CCD data into guide signals will be provided by the French group that is writing the similar control software for the COROT mission.

### 3.2.3 Data Analysis

Image data from GEST will amount to about 170 Gbytes per day when uncompressed. This large data volume will make distribution of a substantial fraction of the image data difficult, so we plan to do the initial data processing at a dedicated data processing facility.

The primary goal of the GBMS analysis is to produce light curves of microlensing events with the greatest possible accuracy and precision. The difference imaging photometry scheme[53,54] has been shown to give optimal performance for ground based microlensing surveys[15,55,55] yielding photometry that is limited primarily by photon noise. This technique involves creating a high spatial resolution, high signal-to-noise reference image which is then transformed to the particular PSF and pointing of the image being reduced. The reference image is then subtracted from the image being reduced, and the resulting difference image is then photometered with a standard aperture or PSF-fitting photometry technique. The difference image contains only positive and negative images of the variable sources in the image, so it is quite sparse even if the original images are quite crowded with stars.

The undersampling problem can be corrected if the data are collected at pointings arranged in a grid-like dither pattern with sub-pixel scale steps. This dithering allows the creation of fully sampled "superimages[25]" which contains all the information necessary to predict the observed image brightness for any sub-pixel pointing offset. This is the basis for the GEST 4×4 grid dither-pattern observing strategy, and the GEST difference imaging scheme will simply replace the undersampled reference image with a reference superimage. A similar approach has been shown to give photon noise limited photometry from images with undersampling and stellar crowding similar to our GBMS observations[30].

A photometry catalog at the data analysis facility will be updated in real time using difference image photometry in conjunction with baseline photometry determined from the reference superimage. To minimize the computational effort, the primary data analysis may be run on a sub-set of the images with photometry on the full light curves being done only for stars that have pass a variation threshold. If the full analysis is run on ¼ of the images, then the total data processed will be only 6 times larger than the data set processed by the MACHO Collaboration a decade before GEST.

PSF fitting photometry on all objects in the master superimage will be performed every month and used to update the baseline magnitude of all objects found in the template image, and the copy of the photometry database at the GEST Science Center at Notre Dame will be updated monthly. Absolute photometry will be reconstructed by cross referencing the baseline photometry with archival HST and ground-based photometry, although the GEST passbands will be wider than standard passbands. Updates to the variable object database at the GEST Science Center will be made continuously for all microlensing events and other unusual variable sources via the Internet (a continuous data stream of < 10 kbytes/sec). Light curve deviations due to exoplanets will result in alerts being sent to the science team and interested follow-up teams as well as being posted on a GEST realtime web-site.

The majority of the GEST observing time will be devoted to the single GBMS field, which will be imaged continuously for 8 months per year. It is, therefore, most efficient to use this same field as a data validation and calibration reference field. The GBMS data analysis routines will incorporate a model of the CCDs, which will be updated when new CCD defects are created by radiation events. The GBMS dithering pattern will allow radiation induced hot pixels to be distinguished from astronomical variability in the GBMS field. The normal GBMS observing program will be modified about once a week for a period of several hours to include a number of dark frames. These will be used to measure the CCD dark current and to detect low-level "warm pixel" defects. A full set of calibration frames will be taken shortly after launch and at the end of each GBMS season. These calibration frames will include long dark exposures, flat field exposures which will be made by rapidly drifting the telescope across regions of the Milky Way and the Earth, and photometric calibration fields which will also serve as CCD cross-talk calibration fields.

## SUMMARY AND CONCLUSIONS

Based on our experience with ground-based gravitational microlensing surveys, we have developed a mission concept capable of detecting 3000 extra-solar planets with high signal-to-noise ratio, including 100-300 Earth-mass planets and 1400 free-floating planets. With a telescope of 1-1.5m aperture and a focal plane array of 32 CCD's, just 3 times as many as were flown on the Chandra ACIS instrument, the GEST mission is feasible with a moderate budget and risk. No other mission concept or ground-based method offers this number of planet detections, especially for low-mass planets like Earth. Follow-up observations with ground-based infrared AO telescopes enable us to observe the host stars and derive the physical parameters of the planetary systems in 1/3 of the cases. We will provide the observational data to test theories of planetary systems and to design new missions capable of collecting planetary photons.

## ACKNOWLEDGMENTS

We would like to thank Bernie Kosicki, Larry Candell, Jim Gregory, Jeff Mendenhall, and Al Pillsbury of MIT's Lincoln Labs and Kin Chan, Paul Forney, Scott Horner, Tom Jamieson, Andy Klavins, and Tom Sherrill for help with the development of the GEST mission concept. Theoretical calculations in support of the GEST mission were supported, in part, by NASA Origins Grants NAG5-4573 and NAG5-9731. Work at the Lawrence Livermore National Laboratory was performed under the auspices of the U.S. Department of Energy, National Nuclear Security Administration under contract No. W-7405-Eng-48.